\documentclass[aps,amssymb,twocolumn,superscriptaddress,unsortedaddress]{revtex4}
\usepackage{setspace}
\usepackage{graphicx}
\usepackage{psfrag}

\def\be{\begin{equation}}
\def\ee{\end{equation}  }
\def\bea{\begin{eqnarray}}
\def\eea{\end{eqnarray}  }

\newcommand{\secintro}{I}
\newcommand{\secoverview}{II}
\newcommand{\secresults}{III}

\newcommand{\secconclusion}{V}

\begin{document}
\title{Evolution of Binary Black Hole Spacetimes}

\author{Frans Pretorius}
\affiliation{Theoretical Astrophysics,
     California Institute of Technology,
     Pasadena, CA 91125}
\affiliation{Department of Physics,
     University of Alberta,
     Edmonton, AB T6G 2J1}

\begin{abstract}
We describe early success in the evolution of binary black
hole spacetimes with a numerical code based on a generalization
of harmonic coordinates. 
Indications are that with sufficient resolution
this scheme is capable of evolving binary systems
for enough time to extract information about the orbit, merger and
gravitational waves emitted during the event. 
As an example we show results from the evolution of a binary 
composed of two equal mass, non-spinning black holes,
through a single plunge-orbit, merger and ring down.
The resultant black hole is estimated to be a Kerr black
hole with angular momentum parameter $a\approx 0.70$. 
At present, lack of resolution far from the binary 
prevents an accurate estimate of the energy emitted, though
a rough calculation suggests on the order of $5\%$ of the initial
rest mass of the system is radiated as gravitational waves
during the final orbit and ringdown.
\end{abstract}

\maketitle


\noindent{\bf{\em \secintro. Introduction:}} One of the more pressing, unsolved problems in general relativity
today is to understand the structure of spacetime describing the
evolution and merger of binary black hole systems. 
Binary black holes are thought to exist in the universe, and the 
gravitational waves emitted during a merger event are expected to 
be one of the most promising sources for detection by 
gravitational wave observatories (LIGO, VIRGO, TAMA, GEO 600, etc.). 
Detection of such an event would be an unprecedented test of general
relativity in the strong-field regime, and could shed light
on many issues related to the formation and evolution of black holes
and their environments within the universe.
Given the design-goal sensitivities of current gravitational 
wave detectors, matched filtering may be essential to detect 
the waves from a merger, and extract information
about the astrophysical source. 
During the early stages of a merger, and the later stages
of the ringdown, perturbative analytic methods should give a
good approximation to the waveform\cite{blanchet,price_pullin};
however, during the last several orbits, plunge, and early stages
of the ring down, it is thought a numerical solution of the 
full problem will be needed to provide an accurate waveform. 

Smarr\cite{smarr} pioneered the numerical study of binary
black hole spacetimes in the mid-70's, where he considered
the head-on collision process in axisymmetry. The full 3D 
problem has, for many reasons, proven to be a more challenging
undertaking, and only recently has progress been made in the 
ability of numerical codes to evolve binary 
systems\cite{bruegmann,brandt_et_al,baker_et_al,bruegmann_et_al,alcubierre_et_al}.
However, until now no code has been able to simulate a non-axisymmetric
collision through coalescence and ringdown. The purpose of this
letter is to report on a recently introduce numerical
method based on generalized harmonic coordinates\cite{paper1} that
{\em can} evolve a binary black hole during these crucial
stages of a merger. At a given resolution the code will not run
``forever'', though convergence tests suggest that with sufficient
resolution the code can evolve the system for as long as needed
to extract the desired physics from the problem. As an example
we describe an evolution that completes approximately
one orbit before coalescence, and runs for long enough 
afterwards to extract a waveform at large distances from the 
black hole. 

The code has several features of note, some or all of which
may be responsible for its stability properties: (1) a formulation
of the field equations based on harmonic coordinates as first
suggested in \cite{garfinkle}, (2) a discretization scheme where
the only evolved quantities are the covariant metric elements, 
harmonic source and matter functions, thus minimizing
the number of constraint equations that need to be solved (which is
similar to the discretization scheme used in \cite{szilagyi_winicour}), (3)
the use of a compactified coordinate system where the outer
boundaries of the grid are at spatial infinity, hence
the physically correct boundary conditions can be placed there, (4)
the use of adaptive mesh refinement to adequately resolve
the relevant length scales in the problem, (5) dynamical excision that
tracks the motion of the black holes through the grid, (6)
addition of numerical dissipation to control 
high-frequency instabilities, (7)
a time slicing that slows down the ``collapse'' of the lapse
that would otherwise occur in pure harmonic time slicing, and (8) the addition
of ``constraint-damping'' terms to the field 
equations\cite{brodbeck_et_al,gundlach_et_al}.
This final element was not present in the version of the code
discussed in \cite{paper1}, and though these terms seem to have
little effect when black holes are not present in the numerical
domain, they have a significant effect on how long a simulation
with black holes can run with reasonable accuracy, at a given
resolution.

An outline of the rest of the paper is as follows.
In Sec. \secoverview\  we give a brief overview of the numerical 
method, focusing on details not present in \cite{paper1}.
Sec. \secresults\  gives results from the simulation of one such orbital
configuration. We conclude
in Sec. \secconclusion\  with a summary of future work.
More details, including convergence 
tests, the effect of constraint damping and a thorough
description of the initial data calculation,
will be presented elsewhere.


\noindent{\bf{\em \secoverview. Overview of the numerical method:}}
We briefly summarize the formulation of the field equations, gauge
conditions and initial data used here, emphasizing details that are not
contained in \cite{paper1}. We discretize the Einstein field equations
expressed in the following form (using units where
Newton's constant and the speed of light are equal to 1): 
\bea
g^{\delta\gamma}g_{\alpha\beta,\gamma\delta}
+ g^{\gamma\delta}{}_{,\beta} g_{\alpha\delta,\gamma}
+ g^{\gamma\delta}{}_{,\alpha} g_{\beta\delta,\gamma}
+ 2 H_{(\alpha,\beta)} \nonumber\\
- 2 H_\delta \Gamma^\delta_{\alpha\beta}
+2 \Gamma^\gamma_{\delta\beta}\Gamma^\delta_{\gamma\alpha}
= - 8\pi\left(2 T_{\alpha\beta}-g_{\alpha\beta} T\right) \nonumber\\
- \kappa\left(n_\alpha C_\beta + n_\beta C_\alpha - 
              g_{\alpha\beta} n^\gamma C_\gamma\right).\label{efe_h}
\eea
$H_\mu$ are (arbitrary) source functions encoding the gauge-freedom of the
solution, $\Gamma^\delta_{\alpha\beta}$ are the Christoffel symbols,
$T_{\alpha\beta}$ is the matter stress tensor with trace $T$, $\kappa$
is a positive constant multiplying the new constraint damping terms 
following\cite{gundlach_et_al}, 
$n^\mu = 1/\alpha (\partial/\partial t)^\mu - \beta^i/\alpha (\partial/\partial x^i)^\mu$
is the unit hypersurface
normal vector with lapse function $\alpha$ and shift vector $\beta^i$ 
($x^0\equiv t$, $x^i \equiv [x^1,x^2,x^3] \equiv [x,y,z]$),
and $C_\mu$ are the constraints:
\be
C_\mu \equiv H_\mu - g_{\mu\nu} \Box x^\nu.\label{c_def}
\ee
We use the following to evolve the source functions:
\be
\Box H_t = - \xi_1 \frac{\alpha-1}{\alpha^\eta} + \xi_2 H_{t,\nu} n^\nu\label{t_gauge}, \ \ \ H_i=0
\ee
where $\xi_1$ and $\eta$ are positive constants.
Note that (\ref{t_gauge}) is 
{\em not} the usual definition of spatial harmonic 
gauge, which is defined in
terms of contravarient components $H^\mu$.

We use scalar field gravitational collapse to prepare initial data 
that will evolve towards a binary black hole system. Specifically,
at $t=0$ we have two Lorentz boosted scalar field profiles,
and choose initial amplitude, separation and boost parameters 
to approximate the kind of orbit that the black 
holes (which form as the scalar field collapses) will have.
The procedure used to calculate the initial geometry is
based on standard techniques\cite{cook_review}, and is
a straight forward extension of the method described
in\cite{paper1} to include non-time-symmetric initial data. 
The initial spatial 
metric and its first time derivative is conformally flat, 
and we specify a slice that is maximal and harmonic.
The Hamiltonian constraint is used to solve for the
conformal factor. The maximal conditions $K=0$ and
$\partial_t K=0$ ($K$ is the trace of the extrinsic curvature)
give the initial time derivative of the conformal factor and
an elliptic equation for the lapse respectively. The
momentum constraints are used to solve for the initial values
of the shift vectors, and the harmonic conditions $H_\mu=0$ are
used to specify the initial first time derivatives
of the lapse and shift. 

\begin{figure}
\begin{center}
\includegraphics[width=6.5cm,clip=false]{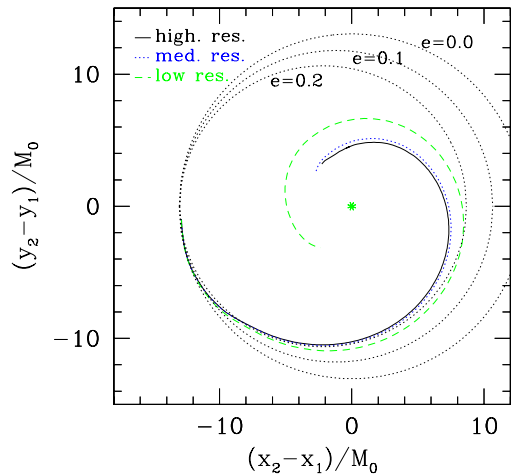}
\end{center}
\caption{A depiction of the orbit for the simulation
described in the text (see also Table \ref{tab_id}).
The figure shows the coordinate position of the center
of one apparent horizon relative to the other, in the 
orbital plane $z=0$. The units have been
scaled to the mass $M_0$ of a single black hole, and
curves are shown from simulations with three different resolutions.
Overlaid on this figure are reference ellipses of eccentricity
$0$, $0.1$ and $0.2$, suggesting that if one were
to attribute an initial eccentricity to the
orbit it could be in the range $0-0.2$.
}
\label{fig_orbit}
\end{figure}

\noindent{\bf{\em \secresults. Results:}}
In this section we describe results from the evolution
of one example of a scalar field constructed binary
system. The present code requires significant
computational resources to evolve binary spacetimes\footnote{Typical 
runtimes on 128 nodes of a Xeon Linux cluster
are on the order of a few days for the lowest resolutions
attempted, to several months at the higher resolutions},
and thus to study the orbital, plunge, and ringdown
phases of a binary system in a reasonable amount of simulation
time we chose initial data parameters such that the
black holes would merge within roughly one orbit---see Fig. \ref{fig_orbit}
and Table \ref{tab_id}.
The following evolution parameters in (\ref{efe_h}) and (\ref{t_gauge}) 
were chosen: 
$\kappa\approx 1.25/M_0,\ \xi_1\approx 19/M_0,\ \xi_2\approx 2.5/M_0,\ \eta=5$
(these parameters did not need to be fine tuned), where
$M_0$ is the mass of one black in the binary. 
This system was evolved using three different grid hierarchies,
which we label as ``low'', ``medium'' and ``high'' resolution.
The low resolution simulation has a base grid
of $32^3$, with up to 7 additional
levels of $2:1$ refinement (giving a resolution in the vicinity
of the black holes of $\approx M_0/10$). For computational efficiency
we only allowed regridding of level $6$ and higher
(at the expense of not being able to accurately
track out-going waves). For the
medium resolution simulation, we have one additional
level of refinement during the inspiral and early phases
of the merger, though have the same resolution
over the coarser levels and at late times; thus we are able to resolve
the initial orbital dynamics more accurately with the medium
compared to low resolution run, though both have roughly
the same accuracy in the wave zone.
The high resolution
simulation has up to 10 levels of refinement during the inspiral
and early ringdown phase, 9 levels subsequently,
and the grid structure of the lower levels is altered
so that there is effectively twice the resolution
in the wave zone. The reason for this set of hierarchies is
again for computational efficiency: doubling (quadrupling) the
resolution throughout the low resolution hierarchy would have
required 16 (256) times the computer time,
which in particular for the higher resolution simulation is 
impractical to do at this stage.

\begin{table}
{\small
\begin{tabular}[t]{| l l || c | c | c |}
\hline
 & & low res. & med. res. & high res.\\
\hline
 & ADM Mass & $2.36 M_0$ & $2.39 M_0$ & $2.39 M_0$ \\
\hline
{\bf initial} & BH masses & $0.97 M_0$ & $0.99 M_0$ & $M_0$ \\
              & orbital eccentricity & $0 - 0.2$ & $0-0.2$ & $0-0.2$ \\
              & proper separation & $16.5 M_0$ & $16.6 M_0$ & $16.6M_0$ \\
              & angular velocity $\times M_0$ & $0.023$ & $0.023$ & $0.023$ \\
\hline
{\bf final} & BH mass & $1.77 M_0$ & $1.85 M_0$ & $1.90M_0$ \\
            & BH spin parameter & $0.74$ & $0.73$ & $0.70$ \\
\hline
\end{tabular}
}
\caption{Some properties of the simulated equal mass binary system 
described in the text. Where relevant the units have been scaled 
to the mass $M_0$ of one of the initial black holes, measured
from the higher resolution simulation at a time after the majority
of scalar field accretion has occurred. The final black hole
mass and spin where estimated from data as shown in 
Fig. \ref{fig_am}, a little while
after the black hole formed, though not so long after
as to be affected by the ``drift'' from numerical error.
The initial proper separation was measured at $t=10 M_0$,
and is the proper length of the piece of a coordinate line
outside the apparent horizons that connects their coordinate
centers. The black holes initially have zero spin.
}
\label{tab_id}
\end{table}

\begin{figure}
\begin{center}
\includegraphics[width=4.1cm,clip=false]{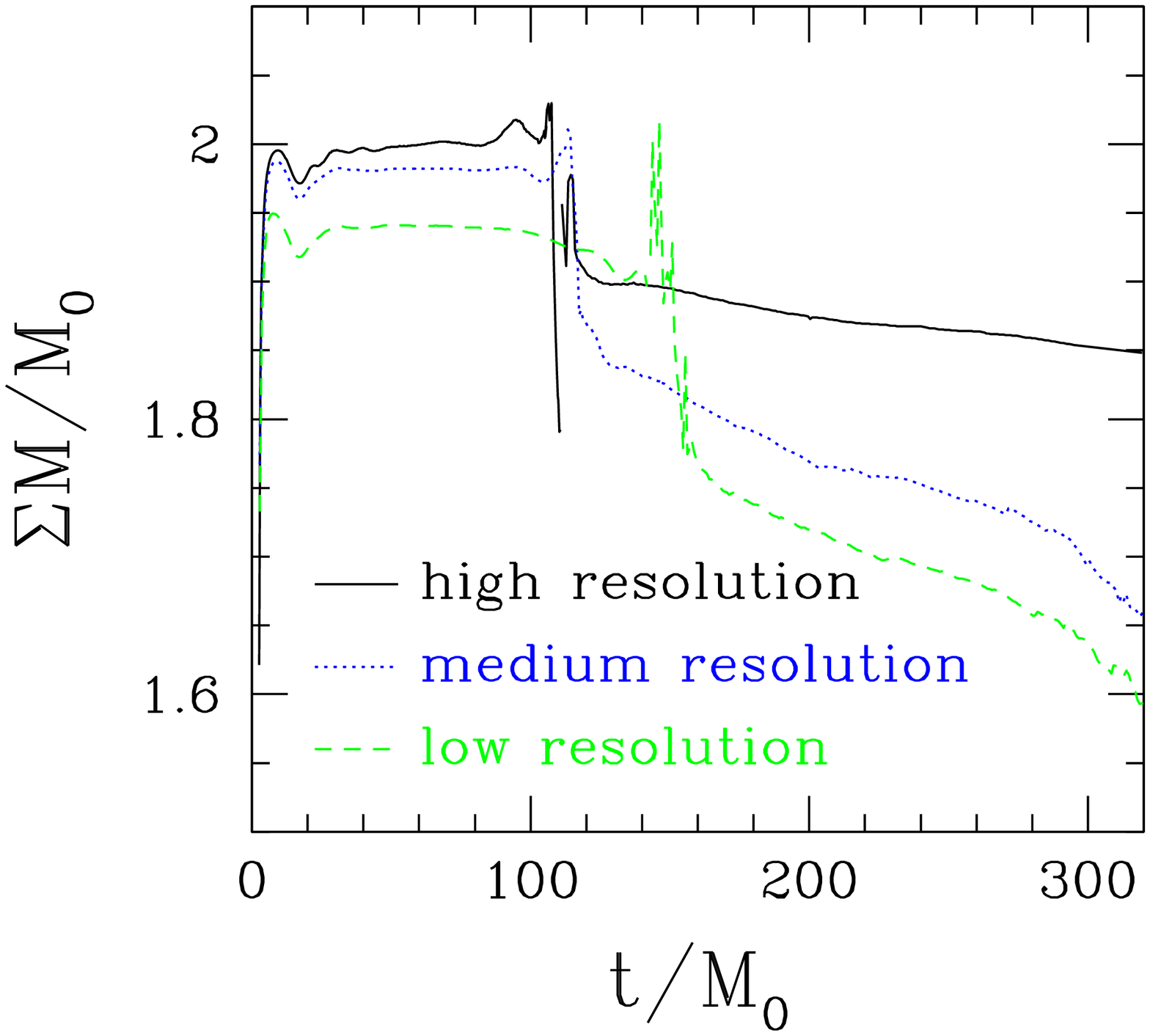}
\includegraphics[width=4.4cm,clip=false]{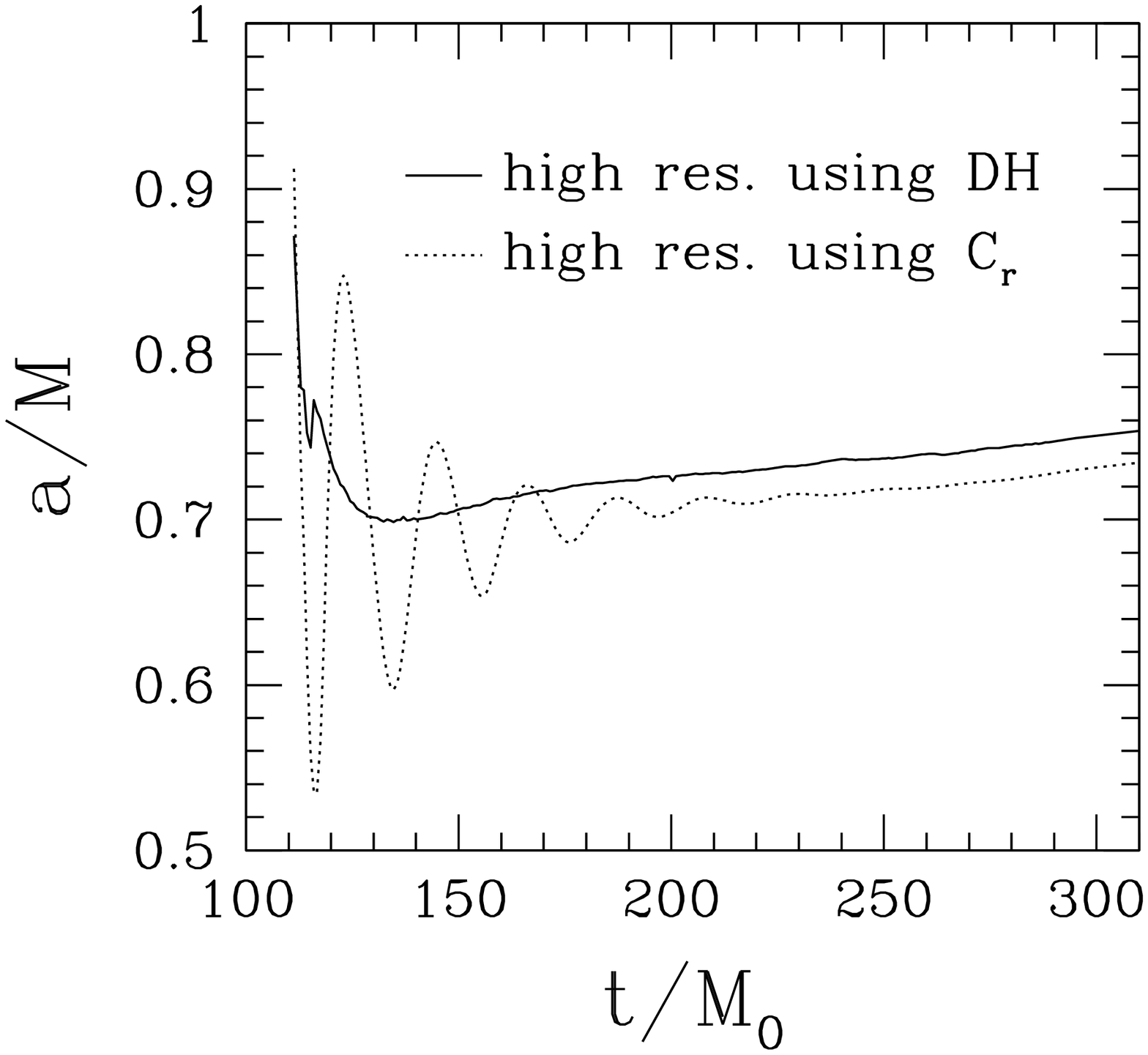}
\end{center}
\caption{The plot to the left shows the net black hole mass 
of the spacetime in
units of the mass $M_0$ of a single initial black
hole, calculated from apparent horizon (AH) properties (using
(\ref{smarr}) with the dynamical horizon estimate
for $J$), and from simulations with three different
resolutions. The initial
sharp increase in mass is due to scalar field accretion,
the small ``wiggle'' at around $20 M_0$ appears
to be a gauge effect, and the ``jaggedness'' around
the time of the merger is due 
to robustness problems in the AH finder that manifest
when the AH shapes are highly distorted.
To the right the Kerr parameter $a$ of the final black hole
is shown (for clarity we only plot the results from
a single simulation), calculated using the ratio $C_r$ of polar
to equatorial proper circumference of
the AH and applying (\ref{a_cr}),
and using the dynamical horizon framework (curve labeled DH).
The loss of mass (and similarly increase in $a$)
with time after the merger is due to
accumulating numerical error.
}
\label{fig_am}
\end{figure}

Fig. \ref{fig_am} shows the horizon masses and final horizon
angular momentum as a function of time.
The ADM mass of the space time
suggests that approximately $15\%$ of the total scalar
field energy does not collapse into black holes.
The remnant scalar field leaves the vicinity of the orbit 
quite rapidly (in $t\approx 30 M_0$, which is on the order of the light
crossing time of the orbit).
Black hole masses are estimated from the horizon area $A$
and angular momentum $J$, and applying the Smarr formula:
\begin{equation}\label{smarr}
M=\sqrt{M_{ir}^2 + J^2/(4 M_{ir}^2)}, \ \ \ M_{ir}\equiv \sqrt{A/16\pi}.
\end{equation}
The horizon angular momentum of the final black hole
is calculated using two methods (which {\em do}
give zero angular momentum when applied to the 
initial black holes, as expected). First, by using the dynamical 
horizon framework\cite{ashtekar_krishnan}, though {\em assuming}
that the rotation axis of the black hole
is orthogonal to the $z=0$ orbital plane, and that each
closed orbit of the azimuthal vector field (which at late times should 
become a Killing vector) lies in a $z={\rm constant}$ 
surface of the simulation. Due to the symmetry of the initial data,
these assumptions are probably valid, though this will eventually need
to be confirmed. The second method, following \cite{brandt_seidel},
is to measure the ratio $C_r$ of the polar to equatorial
proper radius of the horizon, and use the 
formula that closely approximates the function that is
valid for Kerr black holes:
\begin{equation}\label{a_cr}
a \approx \sqrt{1-\left(2.55 C_r -1.55\right)^2}
\end{equation}
As seen in Fig. \ref{fig_am}, the initial ringing of the black hole
is quite apparent in the estimate using $C_r$.
Remarkably, the dynamical horizon estimate for $a$ and 
average value obtained using $C_r$ 
agree quite closely, even shortly after the merger when one
might have expected the black hole to still be too
far from its stationary state to have either method
be applicable.

\begin{figure}
\begin{center}
\includegraphics[width=8.2cm,clip=false]{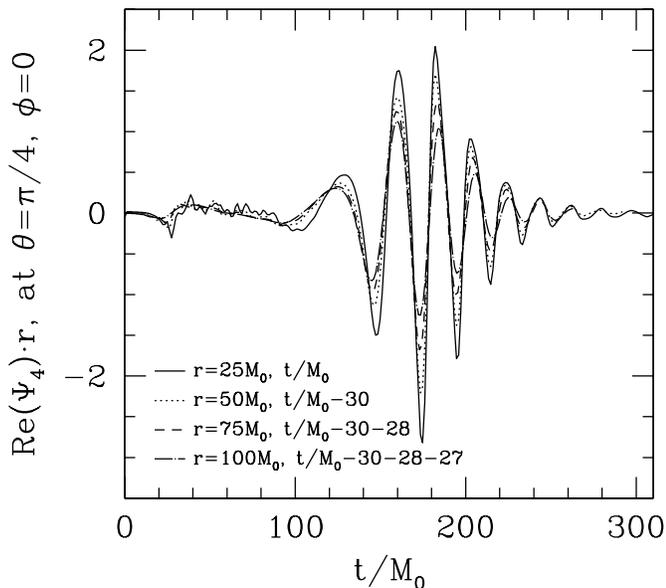}
\end{center}
\caption{ A sample of the gravitational waves emitted
during the merger, as estimated by the Newman-Penrose
scalar $\Psi_4$ (from the medium resolution simulation). 
Here, the real component of $\Psi_4$
multiplied by the coordinate distance $r$ from the center
of the grid is shown at a fixed angular location, though several
distances $r$. The
waveform has also been shifted in time by amounts
shown in the plot, so that the oscillations overlap.
If the waves are measured far enough from the central black hole 
then the amplitudes should match, and they should be shifted
by the light travel time between the locations (i.e. by $25 M_0$ 
in this example). That we need to shift the waveforms by
more than this suggests the extraction points
are still too close to the black hole; the
decrease in amplitude is primarily due to numerical
error as the wave moves into regions of the grid
with relatively low resolution.}
\label{fig_psi4}
\end{figure}

To estimate the gravitational waves emitted by the binary
we use the Newman-Penrose scalar $\Psi_4$, with the null
tetrad constructed from the unit timelike normal $n^\mu$,
a radial unit spacelike vector normal to $r={\rm constant}$
coordinate spheres, and two additional unit spacelike
vectors orthogonal to the radial vector\footnote{At this stage we are ignoring 
all the subtleties
in choosing an ``appropriate'' tetrad}.
Far from the source, the 
real and imaginary components of $\Psi_4$ are 
proportional to the second time derivatives of the
two polarizations of the emitted gravitational
wave.
Fig. \ref{fig_psi4} shows an example of the real part of $\Psi_4$.
Most of the early, short wavelength
burst of waves can be correlated with the passage
of the remnant scalar field that did not fall into the
black holes (the ``noisy'' nature of this piece of the
waveform is in part due to numerical error). This unwanted
radiations leaves the domain quite early on, and so 
does not significantly affect the subsequent merger waves.
Roughly the first long wavelength oscillation in the
plot can be associated with orbital motion, and
subsequent waves with the ringdown of the final black hole.

To estimate the total energy $E$ emitted in gravitational waves,
we use the following formula \cite{smarr_rad}
\begin{equation}
\frac{dE}{dt}=\frac{R^2}{4\pi} \int p d\Omega, \ \ \
p=\int_0^t \Psi_4 dt \cdot \int_0^t \bar{\Psi}_4 dt \label{Ep},
\end{equation}
where $\bar{\Psi}_4$ is the complex conjugate of $\Psi_4$,
and the surface integrated over in (\ref{Ep}) is 
a sphere of constant coordinate radius $R$ (in uncompactified
coordinates). 
This method of calculating the energy is
quite susceptible to numerical error, as we are
summing a positive definite quantity
over all time to give a {\em change} of energy
with respect to time; thus numerical error in $\Psi_4$ will
tend to inflate the answer.
To reduce some of this error, we filter out the high spherical harmonic 
components ($\ge\ell=|m|=6$) of $\Psi_4$ before
applying (\ref{Ep}). 
Note that the smaller integration radii (as shown in Fig. \ref{fig_psi4}) 
are not very far from the binary system, and so possibly
in a region where (\ref{Ep}) is not strictly valid.
However, the larger integration radii are in regions of the 
grid that do not have very good resolution (due both to the 
mesh refinement structure and the spatially compactified 
coordinate domain), and so numerical error (mostly dissipation)
tends to reduce the amplitude of the waves with distance
from the source.
With all these caveats in mind, the numbers we
obtain from (\ref{Ep}) are $4.7\%, 3.2\%, 2.7\%, 2.3\%$
at integration radii of $25M_0, 50M_0, 75M_0$ and $100M_0$
respectively (from the high resolution simulation\footnote{the corresponding
numbers from the medium(low) res. runs are
$5.1(7.1)\%$, $3.5(4.6)\%$, $2.5(3.2)\%$, $1.7(2.1)\%$}), and
where the percentage is relative to $2 M_0$.
Another estimate
of the radiated energy can be obtained by taking
the difference between the final and initial
horizon masses (Table \ref{tab_id})---this suggests
around $5\%$ (high resolution case).


\noindent{\bf{\em \secconclusion. Conclusion:}}
In this letter we have described a numerical method based
on generalized harmonic coordinates that can stably evolve
(at least a class of) binary black hole spacetimes. 
As an example, we presented an evolution of a binary system
composed of non-spinning black holes of equal mass $M_0$,
with an initial proper separation and orbital angular velocity 
of approximately $16.6M_0$ and $0.023/M_0$ respectively.
The binary merged within approximately 1 orbit, leaving
behind a blackhole of mass $M_f \approx 1.9 M_0$ and
angular momentum $J \approx 0.70 M_f^2$. A calculation
of the energy emitted in gravitational waves indicates
that roughly $5\%$ of the initial mass (defined as $2 M_0$)
is radiated . Future work includes improving the accuracy of
simulation (in particular the gravitational waves),
exploring a larger class of initial conditions 
(binaries that are further separated,
have different initial masses, non-zero spins, etc.), and
attempting to extract more geometric information about the nature
of the merger event from the simulations.


\noindent{\bf{\em Acknowledgments:}}
I would like to thank Carsten Gundlach for describing their constraint
damping method for the Z4 system\cite{gundlach_et_al}, and suggesting that 
it can be applied in a similar fashion with the harmonic scheme.
I would also like to thank Matthew Choptuik, Luis Lehner and Lee Lindblom
for stimulating discussions related to this work.
I gratefully acknowledge research support from
NSF PHY-0099568, NSF PHY-0244906 and Caltech's Richard Chase Tolman Fund.
Simulations were performed on UBC's {\bf vn}
cluster, (supported by CFI and BCKDF), and the {\bf Westgrid} cluster
(supported by CFI, ASRI and BCKDF).

\end{document}